\documentclass[aps,pre,twocolumn,groupedaddress,showpacs]{revtex4-1}
\usepackage{amsmath,amssymb,subfigure,fancybox,txfonts,bm,color}
\usepackage[dvipdfmx]{graphicx}
\begin{document}
\title{
Structural relaxation affecting shear transformation avalanches in metallic glasses
}

\def\tr#1{\mathord{\mathopen{{\vphantom{#1}}^t}#1}} 
\newcommand{\vc}{\mathbf}
\newcommand{\del}[2]{\frac{\partial #1}{\partial #2}}
\newcommand{\gvc}[1]{\mbox{\boldmath $#1$}}
\newcommand{\fracd}[2]{\frac{\displaystyle #1}{\displaystyle #2}}
\newcommand{\ave}[1]{\left< #1 \right>}
\newcommand{\red}[1]{\textcolor[named]{Red}{#1}}
\newcommand{\blue}[1]{\textcolor[named]{Blue}{#1}}
\newcommand{\green}[1]{\textcolor[rgb]{0,0.6,0}{#1}}
\newcommand{\dev}[2]{\frac{\text{d} #1}{\text{d}#2}}
\newcommand{\mdev}[3]{\frac{\text{d}^{#3} #1}{\text{d}#2^{#3}}}
\newcommand{\intd}[1]{\text{d} {#1}}

\newcommand{\eng}[1]{#1}
\newcommand{\jpn}[1]{}
\newcommand{\subti}[1]{\begin{itemize} \item {\bf #1} \end{itemize}}
\newcommand{\niyama}[1]{\textcolor[named]{Red}{(新山) #1}}

\author{Tomoaki Niiyama$^{1}$}
\email{niyama@se.kanazawa-u.ac.jp}
\author{Masato Wakeda$^{2}$}
\author{Tomotsugu Shimokawa$^{3}$}
\author{Shigenobu Ogata$^{4, 5}$}

\affiliation{
$^{1}$
College of Science and Engineering, Kanazawa University,
Kakuma-machi, Kanazawa, Ishikawa 920-1192, Japan\\
$^{2}$
Research Center for Structural Materials,
National Institute for Materials Science,
1-2-1 Sengen, Tsukuba, Ibaraki 305-0047, Japan\\
$^{3}$
Faculty of Mechanical Engineering, Kanazawa University,
Kakuma-machi, Kanazawa, Ishikawa 920-1192, Japan\\
$^{4}$
Department of Mechanical Science and Bioengineering, Osaka University,
Osaka 560-8531, Japan\\
$^{5}$
Center for Elements Strategy Initiative for Structural Materials (ESISM),
Kyoto University, Sakyo, Kyoto 606-8501, Japan
}
\date{\today}

\begin{abstract}

Avalanche behaviors, characterized by power-law statistics 
and structural relaxation that induces shear localization 
in amorphous plasticity, play an essential role 
in deciding the mechanical properties of amorphous metallic solids
(i.e., metallic glasses). 
However, their interdependence is still not fully understood.
To investigate the influence of structural relaxation 
on elementary avalanche behavior,
we perform molecular-dynamics simulations for the shear deformation test
of metallic glasses using two typical metallic-glass models comprising 
a less-relaxed (as-quenched) glass and a well-relaxed (well-aged) glass 
exhibiting a relatively homogeneous deformation 
and a shear-band-like heterogeneous deformation, respectively.
The data on elementary avalanches obtained from both glass models 
follow the same power-law statistics with different maximum event sizes,
and the well-relaxed glass shows shear localization.
Evaluating the spatial correlation functions of
the nonaffine squared displacements of atoms during
each elementary avalanche event,
we observe that the shapes of the elementary avalanche regions
in the well-relaxed glasses tend to be anisotropic, whereas those in
the less-relaxed glasses are relatively isotropic.
Furthermore, we demonstrate that a temporal clustering in
the direction of the avalanche propagation emerges, and 
a considerable correlation between the anisotropy and 
avalanche size exists in the well-relaxed glass model.

\end{abstract}
\maketitle

\section{Introduction}

\label{sec:Intro}

Whereas metals usually form ordered (i.e., latticed) structures
under moderate conditions, various glass states can be realized
by quenching the multicomponent metals from their liquid state.
These types of amorphous metals are called {\it metallic glasses} \cite{Masumoto1971MG,Greer1995MG}.
Such materials have excellent material properties, such as high strength,
corrosion resistance, and soft magnetic properties 
\cite{Greer1995MG,Masumoto1971MG,Loffler2003ReviewMG,Ashby2006MGforStructuralMater}.
However, a high level of macroscopic brittleness and catastrophic failure 
of the metallic glasses caused by {\em shear localization} 
hinders their applicability as structural materials, 
whereas localized deformation (i.e., shear banding) 
induces ductility at microscopic scales \cite{Pampillo1972MGshearFracture,Zhang2003MGshearFracture,Ashby2006MGforStructuralMater,Gludovatz2013MGfatigueFracture}. 
Thus, this localization of deformation is a major concern for the brittle failure of metallic glasses.

Some experimental and numerical studies have reported that 
structural relaxation using specific thermal treatments determined 
whether the shear plastic deformation of metallic glasses was localized 
or homogeneous \cite{Shi2007MG-DisorderT,Kumar2009MGembrittle,Ogata2006localizationMG,Zhang2015Processing-dep-MG,Wakeda2015MGRejuv,Miyazaki2016RejuveMG}. 
Studies on molecular dynamics (MD) simulations indicated that 
well-relaxed glasses (i.e., well-aged glasses) using thermal relaxation 
exhibit shear banding by localized deformation, whereas less-relaxed glasses 
(i.e., as-quenched glasses) exhibit homogeneous deformation \cite{Shi2007MG-DisorderT,Zhang2015Processing-dep-MG,Wakeda2015MGRejuv,Miyazaki2016RejuveMG}. 
Thus, understanding the effects of structural relaxation 
by thermal treatments on the shear localization is expected to lead 
to an improvement of the ductility of metallic glasses.

In the context of nonequilibrium physics, the avalanche behavior 
in plasticity (intermittent plasticity or {\em avalanche plasticity}) 
observed in various types of amorphous solids, 
such as glasses, granular materials, colloids, and metallic glasses 
has gathered considerable attention \cite{Dalton2001GranularSOC,Maloney2004AmorphousAvalanches,Salerno2012AmorphousAvalanches,Hatano2015granular-avalache,Sun2010MetallicGlassSOC,Ren2012MG-Chaos2SOC,Antonaglia2014BulkMetallicGlassSOC}. 
One of the significant characteristics of avalanche plasticity 
is its power-law statistics.
The probability that a plastic event of size of $s$ occurs 
is proportional to an algebraic function, $P(s) \propto s^{-\alpha}$, 
where $\alpha$ is a constant. 
This statistical feature, following a power-law distribution, 
is a sign of nonequilibrium critical phenomena, 
including self-organized criticality \cite{Bak1987SOC,SOC1998Jensen}.
The same power-law behavior also emerges in crystalline solids 
through the collective motion of dislocations, 
and the behavior is thought to represent the intrinsic nature of plasticity
in solids \cite{Miguel2001Intermittent-di,Zaiser2006IntermittentPlasticityReview,Csikor2007DislocationAvalanche,Ispanovity2014NotDepinning,Cui2017DDDSOCNano,Papanikolaou2018AvalanchePlasticityOverview}.
The plastic events obeying power-law statistics in amorphous solids 
correspond to avalanche-like collective motions 
of local atomistic rearrangements.
The minimum unit of plastic deformation in amorphous solids 
is considered to be a set of atomic rearrangements 
in a local region known as a shear transformation zone (STZ)
\cite{Argon1979STZ,FalkLanger1998STZ}.
The deformation of an STZ can activate the deformation of other STZs through the redistribution of the elastic energy stored in the STZ.
The chain-reaction propagation of this type of deformation in STZs behaves like an avalanche (we refer to this as an elementary avalanche as described 
in Section ~\ref{sec:avalanche-statistics}). 
The plastic deformation of amorphous solids is a result of 
several shear transformation avalanches
\cite{Boioli2017STZactivationVolume}. 
Thus, the localization of the deformation 
(i.e., shear banding in amorphous solids) can be considered 
a spatial concentration of shear transformation avalanches. 
Hence, avalanche plasticity is expected to be closely related to 
ductility in metallic glasses \cite{Sun2010MetallicGlassSOC}.

Annealing's influence of structural relaxation on avalanche plasticity (e.g., the difference of the avalanche behaviors between the localized deformation of well-relaxed glasses and homogeneous deformation of less-relaxed glasses) is still not well understood.
Furthermore, it is not known how avalanches in well-relaxed glasses result 
in shear banding or how structural relaxation influences 
the shape of the avalanche.
One may expect that less-relaxed glasses would not exhibit avalanche behaviors, because the excess free volume within their atomic configurations should prevent avalanche formation. 
Elucidating the relationship between the localization and behavior of the avalanche plasticity is expected to contribute to the improvement of the mechanical properties of metallic glasses.
Furthermore, it is expected to provide an understanding of nonequilibrium critical behaviors of amorphous plasticity.

In this study, we investigate the influence of structural relaxation 
on avalanche plasticity and the contribution of the avalanche 
to shear localization via MD simulations using well-relaxed and less-relaxed 
metallic glasses exhibiting localized and homogeneous deformations 
induced by specific thermal treatments. 
First, we confirm the localization and the avalanche statistics 
of plastic deformation in the two metallic glasses
(Section ~\ref{sec:avalanche-statistics}). 
Next, we analyze the avalanche shapes in the two metallic glasses, 
(Section ~\ref{sec:avalanche-geometry} and Section ~\ref{sec:spatio-temporal}),
by extracting individual avalanche events from our simulations 
and calculating spatial correlation functions 
of the nonaffine squared displacements \cite{FalkLanger1998STZ}.
Finally, we discuss the evolution of the avalanche shape over time
and the correlation of the shape with the magnitude 
of the avalanche events (Sections ~\ref{sec:anisotropy-size} and ~\ref{sec:size-depend}).

\section{Numerical method}

\label{sec:method}

In the present study, the atomic structure of a less-relaxed glass model with homogeneous deformations was prepared by quenching a copper--zirconium (Cu--Zr) binary alloy from the liquid state, 
whereas a well-relaxed glass model with localized deformations was achieved using specific thermal annealing of the homogeneous glass, as described below.

For these two typical glass configurations, 
we performed MD simulations of shear deformation with constant temperatures and strain rates. 
We selected the Cu--Zr system to perform the simulations, because this alloy exhibits excellent glass formability \cite{Matsubara2007Zr-basedMG}. 
To perform the MD simulations, we used the Lennard--Jones potential and parameters for the Cu--Zr mixtures, which were developed by Kobayashi {\it et al.} \cite{Kobayashi1980LJforCuZrAlloy}. 
The atomic radii of Cu and Zr at this potential are approximately 2.7 and 3.3~\AA, respectively \cite{Kobayashi1980LJforCuZrAlloy}. 
The number of atoms in the simulations was $50,000$ with a $1:1$ ratio 
of Cu and Zr atoms. 

To obtain the atomic structure of the two different glasses,
we applied a specific thermal loading with the following conditions used 
in a previous study to generate either homogeneous or localized deformations 
\cite{Wakeda2015MGRejuv}.
First, the randomly packed configurations of the Cu--Zr atoms 
under the periodic boundary condition were heated to 
a temperature of $3,000$~K, greater than the melting temperature.
Furthermore, the molten configurations were equilibrated for $100$~ps 
under an isothermal-isobaric (NPT) ensemble with zero normal stresses 
after performing equilibration for $100$~ps at the same temperature 
under the canonical ensemble (NVT ensemble).
Quenching the equilibrated liquid to $0$~K with a cooling rate $10^{13}$~K/s 
resulted in the formation of a less-relaxed glass with homogeneous deformation,
because the structure did not undergo any structural relaxation by annealing.
Henceforth, we refer to the glass as the {\it as-quenched model}.

The glass structure exhibiting localized deformation was obtained after additional thermal loading resulting in structural relaxation.
After quenching the equilibrated liquid to $0$~K with 
a relatively slow cooling rate of $10^{12}$~K/s,
we heated the quenched structure to $900$~K (slightly higher than the glass transition temperature \cite{Wakeda2015MGRejuv} $T_g = 898$~K),
then we annealed it for $2$~ns under the NPT ensemble.
Next, we quenched the well-annealed configurations to $0$~K 
at a rate of $3 \times 10^{11}$~K/s.
This thermal annealing leads to a structural relaxation 
in the glass structure without recrystalization,
whereas the annealing temperature was slightly higher than $T_g$.
We confirmed that the radial distribution function of
this well-annealed configuration did not show a significant difference
with the as-quenched configuration.
The structural relaxation resulting from this annealing process
was established by the evaluation of the change
in the atomistic volumes (as illustrated in the next paragraph)
and the aging of the as-quenched and the well-annealed configuration
\cite{Wakeda2015MGRejuv}.
Henceforth, this well-relaxed glass configuration will be referred to 
as the {\it well-aged model}.

The simulation cell volume of the as-quenched model at the initial state 
was $V_{AQ} = 893.97 \pm 0.02$ nm$^3$, 
whereas that of the well-aged model was $V_{WA} = 890.19 \pm 0.02$ nm$^3$.
The variation of the number density was $(\rho_{WA}-\rho_{AQ})/\rho_{AQ} = (V_{AQ}-V_{WA})/V_{WA} \simeq 0.42 \%$, where $\rho_{AQ}$ and $\rho_{WA}$ were the number densities of the as-quenched and the well-aged model, respectively.
This density variation is comparable to those reported in some experimental and numerical studies \cite{Gerling1982MGDensity,Nagel1998FreeVolumeMG,Chen1978MGStructRelax}.
This variation indicates that 
the as-quenched model contained a larger atomic-free volume 
than the well-aged model 
because structural relaxation occurred during thermal annealing
in the latter model.

For these two typical glass configurations, we added simple shear deformation with an engineering strain rate,
$\dot{\gamma} = \dot{\gamma}_{zx} = 10^7$~1/s, for $100$~ns
under the NPT ensemble condition at $10$~K using the Lee--Edwards periodic boundary condition \cite{Tuckerman2010MDsimulations} and zero normal stresses using the Parrinello--Rahman method \cite{Parrinello1980RahmanMethod},
where $\gamma_{xy}$ and $\gamma_{yz}$ were fixed at zero.
The simulation cell for the periodic boundary condition had a cubic shape
with $9.62$~nm edges at the initial state.
The above simulations were performed using LAMMPS \cite{Plimpton1995LAMMPS}.

To remove thermal fluctuations from the original time series
of the shear stress, $\sigma^*_{xz}(t)$, obtained from the simulations,
we smoothed the time series using a Gaussian filter.
Gaussian filtering is an averaging method involving a Gaussian weight.
This is shown in the following equations.
\begin{align}
 \sigma_{xz}(t) = \int^{\infty}_{-\infty} G(t'-t) 
\cdot \sigma^*_{xz}(t) \ \intd{t'},
\end{align}
where $G(t)$ is the Gaussian weight, and
\begin{align}
 G(t) = \frac{1}{\sqrt{2 \pi \delta^2}} \exp \left[ - t^2/2 \delta^2 \right],
\end{align}
where $\delta$ represents the extent of the filter.
In this study, we computed the sum over a discrete range
from $-3 \delta$ to $+3 \delta$, as shown below,
instead of calculating the above integral over an infinite range,
because $\sigma^*_{xz}(t)$ is discrete time series.
\begin{align}
 \sigma_{xz}(m \Delta t) 
= \sum^{m+3d}_{n=m-3 d} \frac{1}{\sqrt{2 \pi d^2}}
\exp \left[ -\frac{(n-m)^2}{2 d^2} \right] \sigma^*_{xz}(n \Delta t),
\end{align}
where $d = \delta/ \Delta t$, $m = t/ \Delta t$.
Here, the standard deviation, $\delta$, and time segment,
$\Delta t$, were chosen as $2$~ps and $4$~fs, respectively.
To quantify the extent of local plastic deformations in the simulations,
we employed the nonaffine squared displacement, $D^2_{\text{min}}$,
developed by Falk and Langer \cite{FalkLanger1998STZ}.
The quantity well-represents atomic displacements that cannot be
represented by affine transformations
(i.e., the atomic displacement by nonelastic deformation in amorphous solids).

\section{Avalanche statistics of as-quenched and well-aged glasses}

\label{sec:avalanche-statistics}

\begin{figure}[tbp]
 \begin{center}
  \includegraphics[width=9cm]{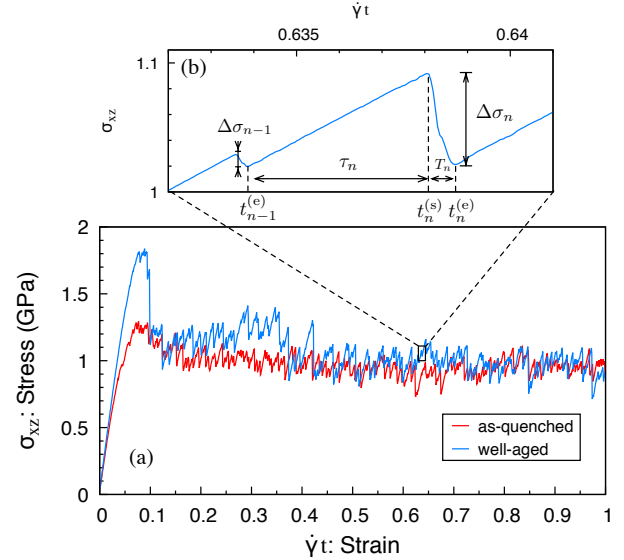}
  \caption{
  (a) Stress-strain curves of the as-quenched model (red)
  and the well-aged model (blue).
  (b) Enlarged view of the stress drops corresponding to the $(n-1)$-th
  and the following $n$-th elementary avalanche events, indicated by the arrows.
}
  \label{fig:t-stress}
 \end{center}
\end{figure}

In Figure ~\ref{fig:t-stress}(a), the evolution over time of the shear stress, $\sigma_{xz}$,
obtained in our MD simulations for the as-quenched and well-aged models,
are depicted by the red and blue lines, respectively.
The stress increased almost monotonously until
$\dot{\gamma} t \simeq 0.05$ ($t \simeq 5$~ns).
Around $\dot{\gamma} t \simeq 0.07$, the well-aged model exhibited a significant overshoot and a sudden descent in shear stress compared to the as-quenched model.
This yielding drop is a common feature of localized deformation of relaxed metallic glasses \cite{Shi2006MGstrainlocalize,Shi2007MG-DisorderT,Zhang2015Processing-dep-MG,Wakeda2015MGRejuv}.
After the drop, both models showed serrated stress fluctuations comprising numerous increases and sudden drops of the shear stress.
Whereas the increasing shear stress was caused by elastic deformation caused by external shear deformation,
the sudden stress drops were the result of plastic deformation (avalanche) events.
This serrated behavior indicates the emergence of avalanche plasticity and is consistent with results obtained from previous experimental and numerical studies of amorphous and crystalline solids \cite{Maloney2004AmorphousAvalanches,Sun2010MetallicGlassSOC,Antonaglia2014BMGSOC,Antonaglia2014BulkMetallicGlassSOC,AnanthakrishnaPRE1999crossover,Niiyama2016avalancheGB}.

%
\begin{figure}[tbp]
 \begin{center}
  \includegraphics[width=9cm]{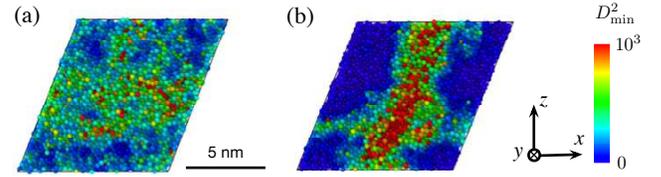}
  \caption{
  Snapshots of the (a) as-quenched and (b) well-aged model
  at $\dot{\gamma} t = 0.4$.
  Atoms are colored according to their nonaffine squared displacements,
  $D^2_{\text{min}}$ (in $\AA^2$), where displacements were calculated
  in reference to the positions of the atoms at $\dot{\gamma} t = 0$.
}
  \label{fig:snapshots-cum}
 \end{center}
\end{figure}

Snapshots of the as-quenched and well-aged models
at $\dot{\gamma} t = 0.4$ are shown in Figs.~\ref{fig:snapshots-cum}(a)
and (b), respectively,
where the color of the atoms represents the magnitude of
the nonaffine squared displacements, $D^2_{\text{min}}$,
from the initial position of each atom \cite{FalkLanger1998STZ}.
The magnitude of the displacement indicates the extent
 of plastic deformation caused by local atomic rearrangements.
Visualization of the snapshots and calculation of the displacements
were performed using the Open Visualization Tool (OVITO) \cite{Stukowski2010Ovito}.
From the snapshots, shear localization occurred in the well-aged model, 
in contrast to the relatively homogeneous deformation pattern 
in the as-quenched model.
Atoms exhibiting plastic deformation localized in two band-like regions along the horizontal and vertical directions in the snapshot of the well-aged model 
[Fig.~\ref{fig:snapshots-cum}(b)].
The emergence of the vertical shear banding was caused by conjugate shear strain, in turn caused by the applied simple shear deformation (a similar behavior of supercooled liquids was discussed in Reference ~\cite{Furukawa2009AnisotropicAmorphousDeform}).
The above results reproducing a heterogeneous deformation 
with the large yielding drop in the thermal annealed metallic glass 
and homogeneous deformation in the non-annealed glass were 
consistent with those of a previous study \cite{Shi2007MG-DisorderT,Zhang2015Processing-dep-MG,Wakeda2015MGRejuv}.

\begin{figure}[tbp]
 \begin{center}
  \includegraphics[width=9cm]{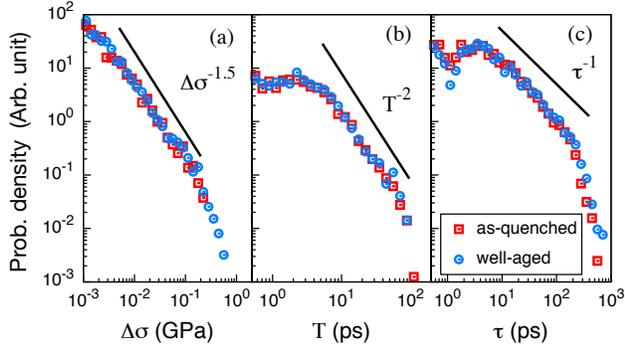}
  \caption{ 
  Statistical distributions of the (a) stress drop, (b) duration,
  and (c) waiting time of plastic deformation (avalanche) events.
  The distributions obtained from the as-quenched and well-aged models
  are represented by red squares and blue circles, respectively.}
  \label{fig:dist}
 \end{center}
\end{figure}

To verify whether our simulations accurately reproduced
the avalanche plasticity,
we calculated the statistical distributions of the stress drop, $\Delta\sigma$, duration, $T$, and waiting time, $\tau$, of each {\em elementary plastic deformation event}.
The elementary plastic deformation event is defined as the plastic deformation caused during a monotonic decrease in the shear stress, $\sigma_{xz}(t)$. The period of the $n$-th deformation event ranges from the start time, $t^{(s)}_{n}$ (where the stress reaches the $n$-th local maximum) to the end time, $t^{(e)}_{n}$ (where the stress reaches the local minimum just after the maximum), as depicted in Fig.~\ref{fig:t-stress}(b).
By the definition, the stress drop, duration, and waiting time of the $n$-th elementary deformation event are determined by $\Delta\sigma_{n} = \sigma_{xz}(t_{n}^{(s)}) - \sigma_{xz}(t_{n}^{(e)})$, $T_{n} = t_{n}^{(e)} - t_{n}^{(s)}$, and $\tau_n = t_{n}^{(s)} - t_{n-1}^{(e)}$, respectively.

The statistical distributions of the stress drop, duration, and waiting time, $P(\Delta\sigma)$, $P(T)$, and $P(\tau)$ are depicted in Figs.~\ref{fig:dist}(a), (b), and (c), respectively.
All distributions decayed algebraically over a range of one or more orders of magnitude (i.e., the stress fluctuation by plastic deformation follows power-law statistics).
The exponents of the power-law distributions are not significantly 
out of the range of those reported in previous studies;
the exponents of event size, duration, and waiting time have been estimated around $1$, $3$, and $1$, respectively, in proceding studies \cite{Maloney2004AmorphousAvalanches,Sun2010MetallicGlassSOC,Hatano2015granular-avalache,Salerno2013inertiaGlassAvalanches,Zhang2017scalingGlassAvalanches}.
This indicates that our simulations successfully reproduced the avalanche behavior of amorphous solids, where each elementary plastic deformation corresponded to one avalanche of plastic deformation.
Thus, we simply refer to elementary deformation events as {\em elementary avalanche events} or {\em avalanche events}.
By comparing the statistics of the as-quenched and well-aged models,
we can see that both distributions follow the same power-law distribution.
However, there is a notable difference in the size distributions,
$P(\Delta \sigma)$, where the maximum size of the stress drops
in the well-aged model is approximately four times larger than that
observed in the as-quenched model [Fig.~\ref{fig:dist}(a)].
Thus, the well-aged glasses have the potential to produce
larger avalanches than the as-quenched glasses,
whereas plastic deformation of the well-aged model is limited to
a narrow band region as depicted in Fig.~\ref{fig:snapshots-cum}(b).

  \section{Technique for evaluating the avalanche geometry}

\label{sec:avalanche-geometry}

In this section, to clarify the spatial features of avalanche plasticity
in the as-quenched and well-aged glasses,
we introduce a spatial correlation function of
the nonaffine squared displacements \cite{FalkLanger1998STZ}
and demonstrate that this correlation function can quantify
the geometry of the elementary avalanche events.
The nonaffine squared displacements for the $n$-th avalanche event
were calculated using the displacements of atoms at $t^{(e)}_n$ with
reference to the positions of the atoms at $t^{(s)}_n$.
We refer to the nonaffine squared displacement of the $i$-th atom
at the $n$-th avalanche event as $D^2_{\text{min}}(\gvc{r}_i(t_n^{(e)}))$,
where $\gvc{r}_i(t_n^{(e)})$ is the position of the $i$-th atom at $t_n^{(e)}$.

\begin{figure}[tbp]
 \begin{center}
  \includegraphics[width=9cm]{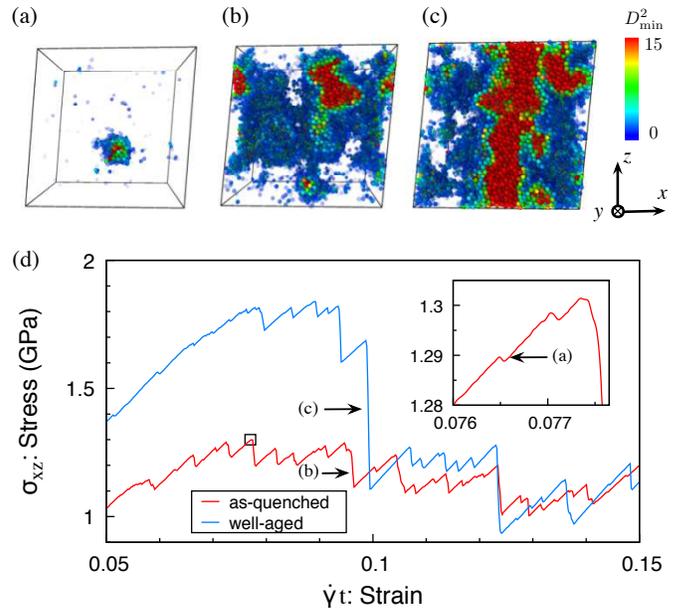}
  \caption{ 
    Snapshots at the three typical elementary avalanche events that
    were associated with stress drops at (a) $\dot{\gamma} t = 0.0765$,
    (b) $0.0960$, and (c) $0.0988$,
    where atoms are colored according to their nonaffine squared displacements,
    $D^2_{\text{min}}(\gvc{r}_i(t_n^{(e)}))$ (in $\AA^2$),
    and only atoms with $D^2_{\text{min}} \ge 1~\AA^2$ are shown (see text).
    (d) Segmentary view of the stress-strain curves including 
    the stress drops corresponding to the snapshots
    (a), (b), and (c) indicated by arrows, respectively.
    Enlarged stress-strain curve of the as-quenched model around 
    the stress drop event at $\dot{\gamma} t = 0.0765$ is imposed.
    }
  \label{fig:snapshots}
 \end{center}
\end{figure}

Typical snapshots of atoms at several avalanche events are depicted 
in Fig.~\ref{fig:snapshots}, where atoms with $D^2_{\text{min}} < 1\ \AA$ 
are not shown.
It can be observed that atoms contributing to plastic deformation tend to localize in space.
As observed in Fig.~\ref{fig:snapshots}(a), the participating atoms clump spherically (the cluster consists of 333 atoms) after a small avalanche with a very small stress drop ($\Delta \sigma = 0.817 \times 10^{-3}$ GPa) occurs in the as-quenched model.
This clumping shape indicates that the avalanche of atomistic rearrangements in the event developed in an {\em isotropic} manner.
In a larger avalanche event that is observed in the as-quenched model 
[Fig.~\ref{fig:snapshots}(b)], the avalanche shape is no longer spherical and does not show any discernible direction.
Thus, it can be considered that this massive avalanche is also isotropic.
This is quantitatively verified later.
In contrast to the as-quenched model, participating atoms in a large avalanche event in the well-aged model clearly gather around a band-like region along the $z$-axis [Fig.~\ref{fig:snapshots}(c)]
(i.e., the avalanche developed {\em anisotropically}).
Note that the highly deformed region (red in the snapshot) approximately corresponds to the shear-banding region depicted {in Fig.~\ref{fig:snapshots-cum}(b).}
Whereas the direct observation is useful to yield intuitive recognition of the geometrical nature of avalanches, the observation is qualitative and depends on the threshold value of $D_{\text{min}}^2$.
Thus, the quantification of the extent of the propagated region in an avalanche event with no consideration of a threshold value is required.
For this, we introduce a planar spatial correlation function, $C_{xy}, C_{yz}$, and $C_{zx}$, of $D^2_{\text{min}}(\gvc{r}_i(t_n^{(e)}))$ at an avalanche event described below.
In this study, $C_{\alpha \beta}$ is introduced as a two-body correlation function that can be given as follows (its correlation is limited to a plane parallel to the $\alpha \beta$ plane):
\begin{align}
  C_{\alpha \beta}(r_{\alpha \beta})
  &=
  \frac{
    \ave{ \delta D(\gvc{r}_i) \cdot \delta D(\gvc{r}_j) }_{ij}^{{\alpha \beta}}
  }
 { \ave{ \delta D(\gvc{r}_i)^2 }_i },
 \label{eq:corr-func}
\end{align}
where $(\alpha, \beta, \gamma)$ are the cyclic permutations of 
$(x, y, z)$, $r_{\alpha \beta}$ is the distance between two points 
on the $\alpha \beta$ plane,
and 
$\delta D(\gvc{r}_i) \equiv D^2_{\text{min}}(\gvc{r}_i(t_n^{(e)})) - \ave{D^2_{\text{min}}(\gvc{r}_k(t_n^{(e)})) }_k$.
The bracket, $\ave{}_k$, refers to the average of any one body scalar quantities
$A(\gvc{r}_k)$ over all the atoms:
$\ave{A(\gvc{r}_k)}_k = \sum_{k=1}^N A(\gvc{r}_k)/N$, 
where $N$ is the total number of atoms in a system.
The bracket, $\ave{}^{{\alpha \beta}}_{ij}$, is the average 
of any two-body scalar quantities, $B(\gvc{r}_i, \gvc{r}_j)$, 
over specific particle pairs, such that the distance
between two atoms on the ${\alpha \beta}$ plane, 
$\sqrt{(\alpha_i-\alpha_j)^2 + (\beta_i-\beta_j)^2}$, 
is in a range from $r_{\alpha \beta}$ to $r_{\alpha \beta} + \Delta r$,
and the distance along the $\gamma$ axis, $|\gamma_i - \gamma_j|$,
is less than $\Delta \gamma$.
We used a $\Delta r$ value of $0.01$~nm 
and selected a sufficiently small width of $0.8$~nm for $\Delta \gamma$.
By employing a window function, the planar average, $\ave{}^{{\alpha \beta}}_{ij}$, is explicitly described as
\begin{align}
 \ave{B(\gvc{r}_i, \gvc{r}_j)}_{ij}^{{\alpha \beta}} &= \frac{  \ave{
 w_{\alpha \beta} ( |\gamma_i-\gamma_j|; \Delta \gamma ) \ w_{r}( r_{\alpha \beta} - |\gvc{r}_{ij}|; \Delta r ) \ B(\gvc{r}_i, \gvc{r}_j)}_{ij} } {\ave{ w_{\alpha \beta} ( |\gamma_i-\gamma_j|; \Delta \gamma ) \ w_{r}( r_{\alpha \beta} - |\gvc{r}_{ij}|; \Delta r )}_{ij} },
\label{eq:ave-ab}
\end{align}
where $w_r$ and $w_{\alpha \beta}$ are the rectangular window functions given as follows:
\begin{align}
  w(x; \Delta x) = \begin{cases}
    1 & ( 0 \le x < \Delta x)\\
    0 & ( \text{otherwise}),
  \end{cases}
\end{align}
and the bracket, $\ave{}_{ij}$, is the average over all particle pairs:
$ \ave{B(\gvc{r}_i, \gvc{r}_j)}_{ij} = \frac{2}{N(N-1)} \sum_{i<j}^N B(\gvc{r}_i, \gvc{r}_j)$.
Note that if one omits $w_{\alpha \beta}$ from Eq. ~(\ref{eq:ave-ab}), $C_{\alpha \beta}$ becomes the conventional spatial correlation function of the distance between any two atoms.
By integrating the correlation function, we can obtain the spatial correlation length of an avalanche event along the $\alpha \beta$ plane corresponding to the average linear size of one avalanche area projected onto the $\alpha \beta$ plane in the following manner:
\begin{align}
 \ave{r}_{\alpha \beta} = { \int r_{\alpha \beta} \ C_{\alpha \beta}(r_{\alpha \beta}) \ \intd{r_{\alpha \beta}} }\  {\LARGE /} { \int C_{\alpha \beta}(r_{\alpha \beta}) \ \intd{r_{\alpha \beta}} }.
\end{align}

The functions, especially correlation lengths, $\ave{r}_{xy}, \ave{r}_{yz}$, and $\ave{r}_{zx}$, indicate how far one avalanche spread parallel to $xy$, $yz$, and $zx$ planar directions, respectively.
The ratio of these correlation lengths provides information regarding the geometry of one avalanche (or the aspect ratio of the avalanche shape).
It should be noted that the spatial correlations introduced in this study were applied to individual avalanche events, whereas similar spatial correlations employed in previous studies were applied to the accumulation of some avalanche events \cite{Nicolas2014GlassSpatioCorr,Shang2014MGstraincorr}.

\begin{figure}[tbp]
 \begin{center}
  \includegraphics[width=9cm]{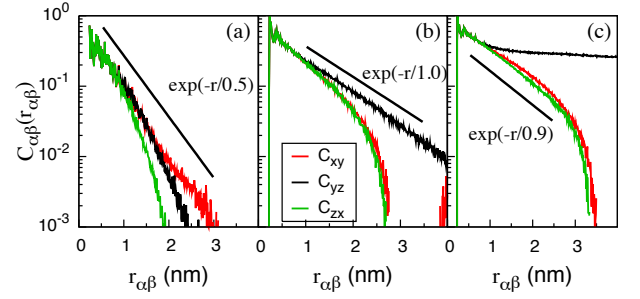}
  \caption{
    Spatial correlation functions of the nonaffine squared displacement
    at $\dot{\gamma} t$ values of (a) $0.0765$, (b) $0.0960$, and (c) $0.0988$,
    where the functions at $\dot{\gamma} t = 0.0765$ and $0.0960$ are
    obtained from the as-quenched model and that at $\dot{\gamma} t = 0.0988$
    is obtained from the well-aged model.
    The correlation functions along the $xy, yz$, and $zx$ planes
    are depicted by the red, blue, and green lines, respectively.
  }
  \label{fig:corr-func}
 \end{center}
\end{figure}

Here we illustrate how this spatial correlation function works to characterize the geometry of the elementary avalanche events using several examples in the as-quenched and well-aged model.
The planar spatial correlation functions for typical avalanches are shown 
in Fig.~\ref{fig:corr-func},
where the correlation functions, $C_{xy}(r_{xy}), C_{yz}(r_{yz})$, and $C_{zx}(r_{zx})$, are colored red, black, and green, respectively.
The correlation functions depicted in Fig.~\ref{fig:corr-func}(a) are for the event at $\dot{\gamma} t = 0.0765$ 
in the as-quenched model.
The corresponding snapshot is shown in Fig.~\ref{fig:snapshots}(a).
All correlation functions in the figure decay following 
the exponential function, 
$C_{\alpha \beta}(r_{\alpha \beta}) \propto \exp[ - r_{\alpha \beta}/\bar{r}]$,
and show the characteristic length, $\bar{r} \simeq 0.5$~nm, 
which is consistent with the average correlation lengths of the event:
$\ave{r}_{xy} = 0.679, \ave{r}_{yz} = 0.725$, and $\ave{r}_{zx} = 0.627$~nm.
The small differences between $\bar{r}$ and $\ave{r}_{\alpha \beta}$ are attributed to the excluded volume effect of atoms that results in a lack of the correlation in the regime, $0 \le r_{\alpha \beta} \lesssim 0.2$~nm.
The result clearly indicates that the avalanche region for this event is almost isotropic, consistent with the direct observation of the event 
[Fig.~\ref{fig:snapshots}(a)].
The correlation functions with larger characteristic lengths,
as shown in Fig.~\ref{fig:corr-func}(b), correspond to 
a larger avalanche event at $\dot{\gamma} t = 0.0960$ 
in the as-quenched model, where the corresponding snapshot is shown
in Fig.~\ref{fig:snapshots}(b).
The correlation functions decay exponentially in accordance with various cut-offs, and
the correlation lengths are $\ave{r}_{xy} = 0.886$, $\ave{r}_{yz} = 1.204$, and $\ave{r}_{zx} = 0.891$~nm.
Whereas these various correlation lengths imply the anisotropic geometry of the event, the difference observed is approximately 36~\% at most.

In contrast with the characteristics of the events in the as-quenched model, a more anisotropic feature can be found in the well-aged model.
The spatial correlation functions of an avalanche at $\dot{\gamma}t = 0.0988$ 
in the well-aged model [Fig.~\ref{fig:corr-func}(c)] show that $C_{yz}$ only shows a small decrease with an increasing $r_{yz}$.
Actually, it follows an algebraic decay, whereas $C_{xy}$ and $C_{zx}$ decay exponentially.
As a result, the correlation length parallel to the $yz$ plane is about twice larger than that to the $xy$ and $xz$ planes.
$\ave{r}_{xy}, \ave{r}_{yz}$, and $\ave{r}_{zx}$ are $1.052, 1.917$, and $0.978$~nm, respectively.
The anisotropy in the estimated correlation lengths of the avalanche is consistent with the snapshot depicted in Fig.~\ref{fig:snapshots}(c).
These results verify that spatial correlation functions and correlation lengths can be used to characterize the geometric features of one avalanche event.
  \section{Spatiotemporal features of anisotropic avalanches}
\label{sec:spatio-temporal}

In this section, we compare overall trends in avalanche geometry
in the as-quenched and well-aged models by estimating the aspect ratio
of all avalanche geometries.
\begin{figure}[tbp]
\centering
 \includegraphics[width=9cm]{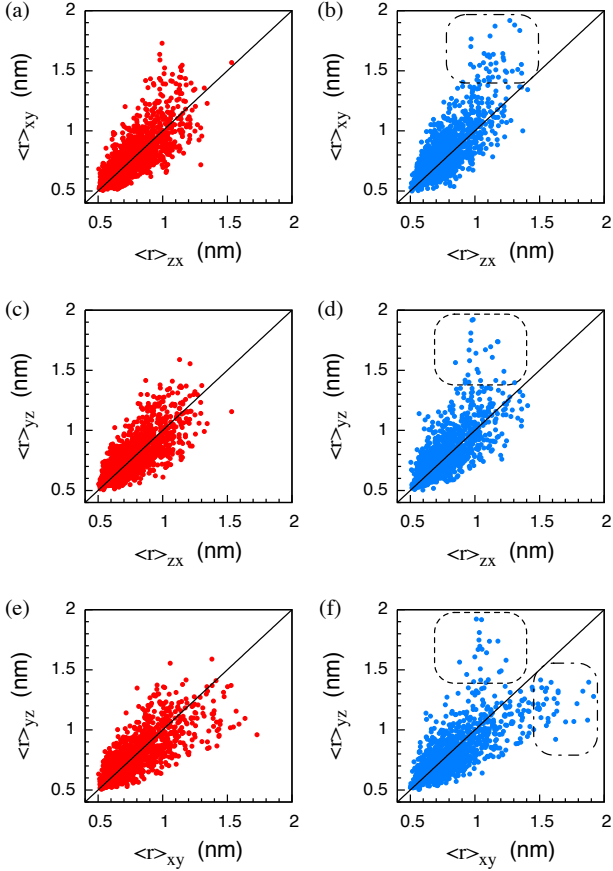}
 \caption{
  Relationship between the spatial correlation lengths of $D^2_{\text{min}}$ 
  on the $xy$, $yz$, and $zx$ planes for (a)(c)(e) the as-quenched model
  and (b)(d)(f) the well-aged model.
}
 \label{fig:ave-R-corr}
\end{figure}

Figure~\ref{fig:ave-R-corr} shows the correlation lengths, 
$\ave{r}_{xy}$, $\ave{r}_{yz}$, and $\ave{r}_{zx}$, 
calculated from all avalanche events 
as functions of each other's correlation length.
The data shown in Figs.~\ref{fig:ave-R-corr}(b) and (d) deviate from the linear relation (indicated by the diagonal line) to the upper-left region of the graphs (i.e., dashed boxes).
This deviation indicates that avalanches in the well-aged model tend to evolve parallel to the $xy$ or $yz$ planar directions rather than the $xz$ direction.
These $xy$ and $yz$ directions correspond to the observed shear banding 
[Fig.~\ref{fig:snapshots-cum}(b)].
In this discussion, we focus on $\ave{r}_{xy}$ and $\ave{r}_{yz}$, the correlation lengths parallel to these two preferable directions.
Figs.~\ref{fig:ave-R-corr}(e) and (f) show $\ave{r}_{yz}$ values obtained in the as-quenched and well-aged models as a function of $\ave{r}_{xy}$.
The data for the well-aged model spread farther away from the diagonal line, compared to those of the as-quenched model at higher correlation lengths indicated by the dashed boxes.
These data indicate that the avalanches in the well-aged model tend to evolve parallel to the two directions, whereas avalanche evolution in the as-quenched model is nearly isotropic.
Thus, structural relaxation by thermal annealing in metallic glasses can enhance both the localization of plastic deformations and anisotropy of the elementary avalanche propagations.

This anisotropy seems to depend on the size of the avalanche event.
In Figs.~\ref{fig:ave-R-corr}(b), (d), and (f), the plots significantly deviate from the diagonal trend line for larger correlation lengths, $\ave{r}_{\alpha \beta} \gtrapprox 1$~nm.
Hence, anisotropy is mainly observed in avalanche events with a large deformation area (as discussed in Section ~\ref{sec:anisotropy-size}).

\begin{figure}[tbp]
\centering
 \includegraphics[width=9cm]{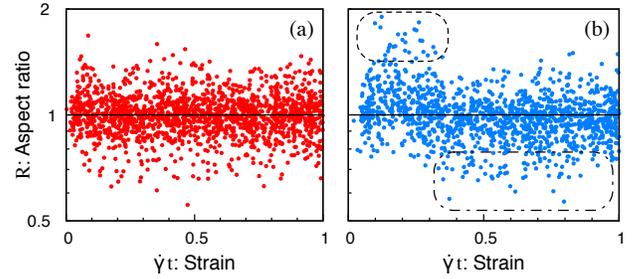}
 \caption{
  Aspect ratios of the region over which the avalanche propagates
  during a plastic event as a function of strain;
  $R = \ave{r}_{yz}/\ave{r}_{xy}$ for the (a) as-quenched model
  and (b) well-aged model.
  The boxes indicated by dashed line and dash-dotted line are events
  during primary and secondary stage, respectively.
}
 \label{fig:t-aspect-R}
\end{figure}
In addition to the spatial correlations, we investigate the temporal clustering of the avalanche anisotropy.
Figure~\ref{fig:t-aspect-R} shows the aspect ratio of the geometry of each avalanche event in the form of the ratio of correlation lengths, $R = \ave{r}_{xy}/\ave{r}_{yz}$, as a function of time (strain).
An $R$ value of 1 for an avalanche signifies isotropic behavior, whereas $R > 1$ and $R<1$ indicate that the anisotropic avalanche propagations are biased toward the $xy$ and $yz$ planar directions, respectively.
Note that, in this figure, the aspect ratio is plotted on a logarithmic scale to show the two directions equivalently.

Figure~\ref{fig:t-aspect-R}(a) shows that $R$ of the as-quenched model is evenly spread around unity, indicating quite isotropic behavior.
This is consistent with the results shown in Figs.~\ref{fig:ave-R-corr}(a), (c), and (e).
In contrast, the aspect ratio of the well-aged model shown in Fig.~\ref{fig:t-aspect-R}(b) tends to be higher than unity in the primary stage (enclosed by the dashed line), whereas in the secondary stage, the $R$ values tend to be lower than unity from around $\dot{\gamma}t = 0.32$ 
(enclosed by a dashed-dotted line).
This indicates that the preferred direction of avalanche propagation in the well-aged model switched from the $yz$ to $xy$ direction at around $\dot{\gamma}t = 0.32$.
The direction of avalanche propagation is not determined randomly but is clustered over time.

This temporal clustering behavior of the avalanche directions in the well-aged model may indicate weakening by plastic deformation.
The region where a previous avalanche occurred can be weakened, 
facilitating subsequent avalanches occurring in the same region and direction.
This behavior implies that avalanche plasticity in amorphous solids 
has the memory of an area that was weakened by previous avalanches.

To comprehensively quantify the tendencies in the anisotropic feature of the avalanche propagation, we calculated the sample deviation, $d$, of the aspect ratio from unity:
\begin{align}
 d^2 = \ave{(R' - 1)^2},
\label{eq:deviation-R}
\end{align}
where the bracket represents the sample average of all avalanche events.
We calculate the aspect ratio, $R'$, as $\ave{r}_{xy}/\ave{r}_{yz}$ or $\ave{r}_{yz}/\ave{r}_{xy}$ in such a way that $R' \ge 1$.
We do this to treat the deviations in the two directions equivalently.
The deviations calculated from the results of the as-quenched model and the well-aged model are $d = 0.1554$ and $0.1890$, respectively.
The deviation of the well-aged model increases by approximately $22$\%.
The result also indicates that the structural relaxation by thermal treatment enhances the anisotropic propagation of the elementary avalanches in metallic glasses.

\section{Correlation between anisotropy and size of an avalanche}

\label{sec:anisotropy-size}

\begin{figure}[tbp]
\centering
 \includegraphics[width=9cm]{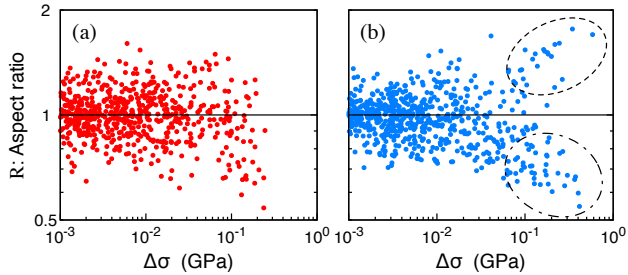}
\caption{
  Relationship between the aspect ratio $R$ and the magnitude 
  of an avalanche event $\Delta \sigma$ for the (a) as-quenched model
  and (b) well-aged model.
}
 \label{fig:aspect-R-ds}
\end{figure}

The relationship between the magnitude and anisotropy of an avalanche event is investigated.
Fig.~\ref{fig:aspect-R-ds} shows the relationship between the aspect ratio, $R$, and the stress drop, $\Delta \sigma$, corresponding to the magnitude of each elementary avalanche event, where only the region of interest of $\Delta \sigma \ge 10^{-3}$~GPa is shown.
It can be seen that $\Delta \sigma$ has a wide range of magnitudes.
The plots of the as-quenched model [Fig.~\ref{fig:aspect-R-ds}(a)] 
show no significant correlation between $\Delta \sigma$ and $R$, whereas there is a clear correlation in the regime $\Delta \sigma > 10^{-1}$~GPa for the well-aged model as shown by the bifurcated plots spreading along the diagonal lines in
Fig.~\ref{fig:aspect-R-ds}(b).
Thus, larger avalanche events result in stronger anisotropy of the deformation region.
This trend is consistent with anisotropic avalanches being mainly observed for large deformation regions, as discussed in Section~\ref{sec:spatio-temporal}.
See also Fig.~\ref{fig:ave-R-corr}(b), (d), and (f).
Thus, elementary deformation avalanches in well-aged metallic glasses
have a tendency to propagate anisotropically in large deformation areas.

The observation that larger avalanches show stronger anisotropy provides interesting insights.
An avalanche of plastic deformation of metallic glasses develops mainly via propagation of local atomistic rearrangements from a small starting region.
Thus, small avalanches can be considered as the propagation of atomic rearrangements with their development halts at an early stage.
In contrast, the development of large avalanches does not stop until a later stage.
Thus, we can consider that a small isotropic avalanche will develop anisotropically along a favored direction from a certain point in time rather than a small anisotropic avalanche increases in size with no change in its aspect ratio.
In other words, the propagation of a deformation avalanche in metallic glasses may bifurcate during the avalanche growth.
The time evolution behavior of the avalanches requires more detailed analysis.

\section{Size dependence of avalanche anisotropy}

\label{sec:size-depend}
\begin{figure}[tbp]
 \begin{center}
  \includegraphics[width=9cm]{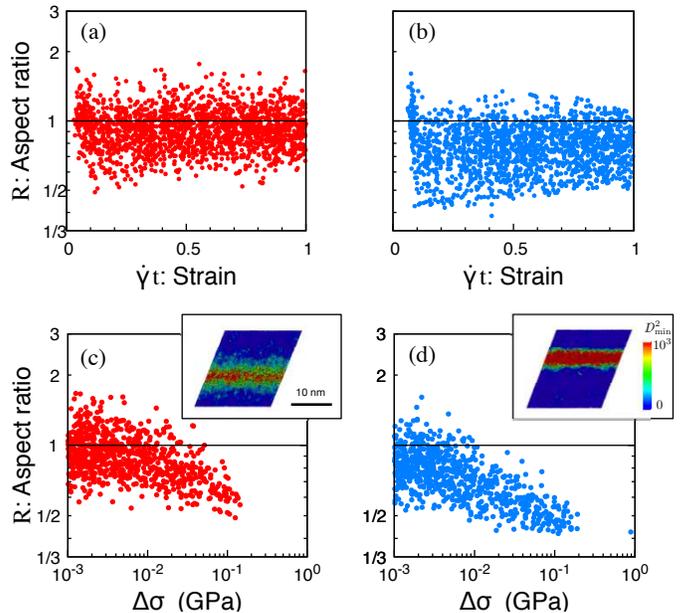}
  \caption{
    (Upper panels) Aspect ratios of the avalanche regions 
    as a function of strain; $R = \ave{r}_{yz}/\ave{r}_{xy}$
    for the larger (a) as-quenched model and (b) well-aged model.
    (Lower panels) Relationship between the aspect ratio $R$
    and the magnitude of an avalanche event $\Delta \sigma$ for
    the larger (a) as-quenched model and (b) well-aged model.
    Snapshots of the larger as-quenched and larger well-aged model
    at $\dot{\gamma} t = 0.4$ are inserted in (c) and (d), respectively.
    Atoms are colored according to $D^2_{\text{min}}$ (in $\AA^2$),
    where the displacements were calculated in reference
    to the positions of the atoms at $\dot{\gamma} t = 0$.
  }
  \label{fig:N=40E+4}
 \end{center}
\end{figure}
Whereas the avalanches in the well-aged model tend to develop anisotropically,
the aspect ratios of these anisotropic avalanches are not as large as
one might expect.
The observed $R$ values remain in the range of $0.5$--$2$, 
as shown in Fig.~\ref{fig:t-aspect-R}(b).
These small values are caused by the finite size effect.
Thus, the integration range for calculating $\ave{r}_{\alpha \beta}$ is limited to half of the length of the simulation cell defined by the periodic boundary condition.
This also explains why $\ave{r}_{yz}$ was only $< 2$~nm, even when $C_{yz}$ decayed algebraically [Fig.~\ref{fig:ave-R-corr}(c)].
In the present simulations, we used a periodic cell with sides of $\sim 9.5$~nm.
Therefore, the correlation length, $\ave{r}_{\alpha \beta}$, can be only $2$~nm at a maximum, even when $C_{\alpha \beta}$ does not decay at all, because the maximum integration range is about $4$~nm.
Thus, MD simulations with larger systems are expected to result in enhanced anisotropic behavior.
To confirm this, we calculated the aspect ratios for additional MD simulations employing larger models with $19.3$~nm edges in the initial state ($400,000$ atoms), performed under the same simulation conditions.
The temporal evolution of the aspect ratios of the elementary avalanche regions and the relationship between aspect ratio $R$ and the event size, $\Delta \sigma$, of each avalanche are shown in Fig.~\ref{fig:N=40E+4}.
Snapshots of the larger as-quenched and well-aged models at $\dot{\gamma} t = 0.4$ are inserted in Figs.~\ref{fig:N=40E+4}(c) and (d), respectively.

The snapshot shows that the larger well-aged model also exhibits shear localization.
An intense concentration of plastic deformation appears in a horizontal band-like area in the larger well-aged model [Fig.~\ref{fig:N=40E+4}(d)].
Whereas the larger as-quenched model also exhibits 
a certain level of localization [Fig.~\ref{fig:N=40E+4}(c)],
its deformation region spreads much wider than that of the well-aged model.
This difference indicates that the structural relaxation caused by thermal annealing  enhances the localization of deformation, even in larger models.

A relatively strong bias in the aspect ratio of isotropic shapes in the larger well-aged model, compared to the one in the larger as-quenched model, is clearly visible in Figs.~\ref{fig:N=40E+4}(b).
The aspect ratios of some extreme events in the larger well-aged model 
drop to $1/3$, whereas a switch in the preferred avalanche direction 
is not observed in this case.
This significant anisotropic trend is also verified by the deviation of aspect ratios defined by Eq. ~(\ref{eq:deviation-R}) in the well-aged model, as $d = 0.473$.
The as-quenched model also exhibits some anisotropic avalanche propagations.
However, the deviation is only $d = 0.259$.
Therefore, the structural relaxation increases the deviation by approximately $83$\%.
As shown in Figs.~\ref{fig:N=40E+4}(c) and (d), whereas the correlation between the magnitude and anisotropy of the avalanche events can be seen in the larger as-quenched model, and not just in the larger well-aged model, the latter is linked to a more definite correlation than the former.
This correlation is more apparent in the larger well-aged model than in the smaller one.
Therefore, it is confirmed that the influence of the structural relaxation on the anisotropy remains, even in the larger systems.
Moreover, the relative influence in the well-aged model, compared to the as-quenched model, grows as the system size increases.

\section{Discussion}

\label{sec:discussion}

The avalanche motion of deformation in amorphous solids can be interpreted as the chain reaction of local shear transformation.
Atoms in a small local area (an STZ \cite{Argon1979STZ,FalkLanger1998STZ}) can rearrange to release external shear stress.
This rearrangement (i.e., the shear transformation of the local atomic configuration) causes stress concentration around the area, as described by the Eshelby inclusion theory, and can induce other local rearrangements through the stress concentration.
The chain reaction of this shear transformation results in a shear transformation avalanche.
Thus, the avalanche behavior illustrated in this study poses 
an important question regarding the manner in which thermal preparation
enhances or suppresses the chain reaction of the shear transformation.

The thermal annealing used in this study causes relaxation of
the internal structural states characterized by various features,
such as short and medium-range order, free volumes, 
or effective disorder temperature of glasses \cite{Sheng2006SRO-MRO-Glass,Cohen1959MolecularTransportLiquid,Shi2007MG-DisorderT,Falk2011STZtheory}.
The structural states and the resultant shear banding
are well described by the effective temperature 
\cite{Shi2007MG-DisorderT,Falk2011STZtheory},
but we herein provide a speculative explanation of the avalanche features 
based on the atomistic free volumes
\footnote{
In this discussion, we consider the atomic free volume not as 
a concept based on the configurational entropy 
but as a primitive one, i.e.,
excess volume per atom derived from a densely packed structure
}
because free volume is a convenient and intuitive concept
in the discussion of atomic-scale dynamics.

As discussed before, 
the as-quenched model exhibiting relatively small avalanches 
has more free volume than 
the well-aged model as mentioned in Sec.~\ref{sec:method},
one can consider that the excess free volume in the as-quenched metallic glasses
could absorb and thus prevent chain reactions in local shear transformations.
Moreover, large free volumes might allow 
the surrounding regions to transform along with directions 
other than that of external shear direction 
because excess free volumes can provide additional space 
for surrounding atoms to move and rearrange themselves, independent 
of the external shear direction.
Conversely, the well-aged glass with less free volume lacks adequate space to absorb the chain reaction.
This may explain the as-quenched glass model neither showing
large nor anisotropic avalanches
\footnote{
The term, ``anisotropy,'' used in this study, refers to avalanche propagation that is dependent on the external loading directions.
Note that this behavior can be confirmed, even in the as-quenched model, as observed in Figs.~\ref{fig:ave-R-corr}(a) and \ref{fig:N=40E+4}(a)
although it is relatively weak.
Thus, the anisotropy discussed in this study should be considered a relative property, depending on the extent of structural relaxation}
.

Although there seems to be a significant relationship between shear banding
and the avalanche behaviors, further researches on this subject 
is essential so that a more comprehensive study based on evidences 
can be performed. 
The scope of the present study comprises, only of 
a comparison of shear-banding patterns [Fig.~\ref{fig:snapshots-cum}(b)]
and the snapshot of a large avalanche event emerging in the well-aged model
[Fig.~\ref{fig:snapshots}(c)];
the focal region of the avalanche 
largely overlaps with the shear banding region.
This shows that relatively few extremely large 
and anisotropic (strip-shaped) avalanches, 
not many small and isotropic avalanches contribute to shear banding 
in metallic glasses.

The influence of environmental temperature on avalanche behaviors,
playing an essential role in amorphous plasticity \cite{Rodney2009distSTZact,Cao2013Strain-T-dependAmorphous,Fan2014STZactivation} is another subject 
of future research.
It is expected that the thermal fluctuation by the temperature will 
provide both enhancement and suppression effects on the avalanche growth and anisotropy;
the enhancement is due to the reduction of atomistic free volume 
caused by structural relaxation driven by the temperature,
and the suppression effect is produced by the thermal activation of the STZs, 
which is supposed to be activated as part of an avalanche 
when the temperature was low, prior to the avalanche.
The opposition of these two effects at high temperature
could affect the avalnache behaviors.

In this discussion, we assumed that the excess free volumes
of metallic glasses leads to the ease of deformation in local area.
This assumption is inferred from the fact that
the as-quench glass with a large free volume 
started to deform at a relatively lower stress than the well-aged glass
with a higher energy state \cite{Wakeda2015MGRejuv} 
as shown in Fig.~\ref{fig:t-stress}.
(Note that a large free volume does not always lead 
to less strength and high energy state \cite{Miyazaki2016RejuveMG}.)
A more valid and comprehensive explanation for the avalanche behaviors
with respect to other factors such as effective temperature 
is necessary and should be a subject of future investigation.

\section{Conclusion}

Using three-dimensional molecular dynamics simulations of shear deformations
in Cu--Zr metallic-glass models,
we compared the avalanche plasticity of a well-relaxed metallic-glass model
(i.e., well-aged model) with localized deformation produced
by thermal annealing and a less-relaxed glass model (as-quenched model)
that did not undergo structural relaxation by annealing,
showing homogeneous deformation.
We focused on analyzing the geometrical feature of elementary shear deformations (i.e., elementary shear transformation avalanches).
The simulations showed that the statistics of stress drops, durations, and waiting times of elementary avalanche events followed power-law distributions for both models.
 The as-quenched model showed a homogeneous deformation pattern, whereas a heterogeneous pattern like a shear band was observed for the well-aged model, as shown in the previous numerical and experimental studies \cite{Shi2007MG-DisorderT,Zhang2015Processing-dep-MG,Wakeda2015MGRejuv}.
We quantified the geometrical features of elementary shear transformation avalanches by introducing a planar spatial correlation function of the nonaffine squared displacements of all atoms in each avalanche event.
The spatial correlations and their characteristic lengths parallel to the $xy, yz$, and $zx$ planes revealed that deformation regions caused by each avalanche event in the well-aged glass tended to be anisotropic, whereas those in the as-quenched glass were generally isotropic.
The direction of anisotropic avalanche propagation in the well-aged glass was not random but showed temporal clustering.
Moreover, we demonstrated that the well-aged glass had a significant correlation between the anisotropy and avalanche magnitude.
The observed differences between the two glass models might be attributed to differences in the atomic-free volume, which can be removed by thermal structural relaxation.

\begin{acknowledgments}
This research was supported by the Ministry of Education, Culture, Sports, Science and Technology (MEXT) KAKENHI Grant-in-Aid for Young Scientists (B) (No. 16K17764), Grant-in-Aid for Scientific Research (A) (No. 17H01238), Grant-in-Aid for Young Scientists (A) (No. 17H04949), Challenging Research (Pioneering) (No. 17K18827).
This research was also supported by 
 ``Exploratory Challenge on Post-K computer'' (Challenge of Basic Science
--- Exploring Extremes through Multi-Physics and Multi-Scale Simulations)
and used computational resources of the K computer 
provided by the RIKEN Advanced Institute for Computational 
Science through the HPCI System Research project 
(Project ID: hp180224 and hp190194). 
\end{acknowledgments}

\bibliographystyle{apsrev}
\bibliography{/Users/niyama/Documents/Articles}

\end{document}